\documentclass[twocolumn,runningheads]{svjour2}
\smartqed  
\usepackage{graphicx}
\journalname{Astrophysics and Space Science}
\begin{document}

\title{Gamma-ray source stacking analysis at low galactic latitudes }

\author{Anal\'{\i}a N. Cillis \and
        Olaf Reimer \and
        Diego F. Torres.
}

\institute{A. Cillis \at
              Goddard Space Flight Center/NASA\\
              Greenbelt 20771, MD, USA\\
              Tel.: +1-301-286-0977 Fax:  +1-301-286-1215\\
              \email{cillis@gamma.gsfc.nasa.gov}           
           \and
           O. Reimer \at
              W.W. Hansen Experimental Physics Laboratory \& \\
	      Kavli Institute for Particle Astrophysics and Cosmology \\
              Stanford University \\
              Stanford, CA 94305-4085, USA\\
              Tel.: +1-650-724-6819 Fax: +1-650-725-2463  \\
             \email{olr@stanford.edu }         
           \and
		   D. Torres \at
		   Instituci\'o de Recerca i Estudis Avan\c{c}ats (ICREA) \& \\
           Institut de Ci\`encies de l'Espai (IEEC-CSIC)\\
           Facultat de Ciencies, Universitat Aut\`onoma de Barcelona\\
           Torre C5 Parell, 2a planta\\ 
		   08193 Barcelona, Spain\\
		   Tel.: +34-93-581-4352 Fax: +34-93-581-4363  \\
             \email{dtorres@ieec.uab.es} 
}
\date{Received: date / Accepted: date}

\maketitle

\begin{abstract}

We studied the problematic of uncertainties in the diffuse gamma radiation apparent in stacking analysis 
of EGRET data at low Galactic latitudes. Subsequently, we co-added maps of counts, exposure and diffuse
 background, and residuals, in varying numbers for different sub-categories of putatively and known 
source populations (like PSRs). Finally we tested for gamma-ray excess emission in those maps and 
attempt to quantify the systematic biases in such approach. Such kind of an analysis will help 
the classification processes of sources and source populations in the GLAST era.
\keywords{gamma rays \and observations \and pulsars \and methods: data analysis}
\PACS{95.85.Pw  \and 98.70.Rz \and 97.60.Gb \and 95.75.-z}
\end{abstract}
\section{Introduction}
The EGRET era fades away while the Large Area Telescope onboard GLAST
 is in its final stages of hardware integration. Many unidentified 
gamma-ray sources are still unidentified, especially at low galactic 
latitudes, where, for instance, not a single supernova remnant could 
be unambigously  detected (e.g., \cite{Torres2003}). Stacking 
techniques are then a powerful tool to explore, in conjunction with 
spatial cross-localization of sources followed up by Monte Carlo 
analysis (e.g., \cite{RBT}), if populations as a whole arise 
in the data. The difficulty being, of course, the correct handling of 
the background gamma-ray emission, which represent a more difficult 
problematic the higher it is. This contribution focuses 
on this very point: devising the first ideas for a method of stacking 
at low Galactic latitudes which not only may allow studying already 
obtained data (i.e., EGRET), but also be applied to the forthcoming 
observations.

In the section below we describe the general ideas of  the stacking 
technique performed. 
In section 3 we comment on the application of this technique to source
population studies in the Galactic Plane (GP),
specifically in the study of pulsars. Our results are given in section
4. To verify the implication of them,we performed simulations, described in 
section 5. Finally, a short discussion is given in section 6. 
\label{intro}
\section{The stacking technique}
The general stacking method we have applied follows that outlined of
reference \cite{CHB}. In order to perform the stacking technique and look for 
a possible collective detection of gamma-ray emission above 100 MeV near the 
GP, we have extracted rectangular sky maps with the selected target 
objects located at the center. We have used EGRET data from April 1991 through 
September 1995 ---matching the baseline of the Third EGRET Catalog \cite{Hartman}, in galactic
coordinates. The extracted maps for each particular target were
chosen to be $60 \times 60$ in size, in order to have large fields of
views and be consistent with the EGRET point spread function (PSF). 
We have transformed the coordinates of each map into pseudo-coordinates, with 
the target object at the center. After doing this, the maps were co-added, 
producing the stacking. It was also necessary to extract a diffuse
background map for each target object. For this purpose, we have used
the diffuse model that is standard in EGRET analysis \cite{Hunter}. 
In order to take into account the existence of identified EGRET sources 
\cite{Hartman,MHR}, idealized sources with the appropriate fluxes distributed 
following EGRET's PSF as well as the modulated artifacts were added to the 
diffuse background. On the other hand, the unidentified EGRET sources were 
not added to the background in this study. It was necessary to normalize each one of the
extracted diffuse maps ($D_i$) for the different exposures ($\epsilon_i$) of the
target objects. The extracted diffuse map for each target object was also transformed 
into pseudo-coordinates. Finally the diffuse maps for the co-added data were obtained as:
$1/\epsilon_{total}\sum_i c_i$ where $c_i$  are the counts
diffuse maps ($c_i=\epsilon_i D_i$) and $\epsilon_{total}=\sum_i \epsilon_i$.
To analyze EGRET data we used the standard likelihood technique 
based upon gamma ray counts maps that were binned in measured gamma-ray energy and 
spatially in rectangular projection in Galactic or celestial coordinates \cite{Mattox}.
The likelihood function of the EGRET data is the probability of the
observed EGRET data for a specific model of high energy gamma-ray emission, and
could be written as the product of the probability for each pixel:
$L_{\theta}=\Pi_j p_j$, where $p_j$ is the Poisson probability of observing $n_j$
counts in pixel $j$ when the number of counts predicted  by the model is $\theta_j$.
The logarithm of the likelihood is used in hypothesis testing and is
usually more easily calculated. Neglecting the last term (model
independent) the logarithm of the likelihood is given by:
\begin{equation}
log L_{\theta}= \sum_j[n_jlog(\theta_j)- \theta_j ].
\label{logL}
\end{equation}
The point-source component of the model consists of an ''active" source
($c_a$ counts
located at ($\alpha_a$,$\delta_a$)) subject to parameter estimation,
and ''inactive" sources with fixed counts at fixed positions. Thus, the total model 
prediction for pixel $j$ is given by
\begin{equation}
\begin{array}{ll}
\theta_j = & g_{mul} G_j + g_{bias}  10^{-5}  E_j + \\
          &+ c_a PSF(\alpha_a,\delta_a, j) +\sum_k
PSF(\alpha_k,\delta_k, j),
\end{array}
\label{theta}
\end{equation}
where $c_k$ is the number of counts for the ''inactive" source at
($\alpha_k$,$\delta_k$); PSF($\alpha$,$\delta$,j) is the fraction of the PSF located at
($\alpha$,$\delta$) that is in pixel $j$; $E_j$  is the exposure in
pixel $j$; and $G_j =\sum_k G_k PSF(\phi_{jk})/\sum_k PSF(\phi_{jk})$
(where ($\phi_{jk}$ is the angle between pixels $j$ and $k$).
The parameters of the gamma-ray model are estimated via the Maximum
likelihood approach. The sum in equation (\ref{logL}) is done for pixels within an
adjustable analysis radius (nominally $15^{\circ}$ for $E>100$ MeV). Within this circle, the
Galactic diffuse radiation model, is scaled by a multiplier, $g_{mul}$, which is estimated by
maximum likelihood. Also, a maximum likelihood value is used for the level of isotropic
diffuse intensity, $g_{bias}$. This decouples the likelihood point-source analysis from 
uncertainties in the large-scale diffuse emision model for analyzing point souces in the 
given region of interest. Only the shape of the model over the
$15^{\circ}$ radius circle is used for point-source analysis. The expected value of $g_{mul}$ is 1 and gbias is 0 if the galactic diffuse model is correct.
In order to test the significance of a detection, the model of
equation (\ref{theta}) is used in the likelihood ratio test by testing the null hypothesis,
$c_a=0$, against the hypothesis that $c_a$ has the estimated value where $g_{mul}$ and
$g_{bias}$ have their optimal values for both hypotheses. This formalism produces a "test statistic'':
$TS=-2(ln L_0 - ln L_1)$, where $L_1$ and $L_0$ are likelihood values with and
without a possible source. $TS^{1/2}$ is roughly equivalent to the standard
deviations.
\label{stacking}
\section{Stacking technique applied to Galactic Plane source population}
The stacking technique for population studies of  gamma-ray sources
has been applied for different class of object in several works (for example:
radiogalaxies and Seyfert galaxies \cite{CHB}, LIRGs and ULIRGs galaxies \cite{CTR}, 
clusters of galaxies \cite{RPSM}).
All these studies have been so far applied only to high-latitute source populations because of the odds 
to deal with a dominant and structured diffuse emission. Here, however, we directly step 
into a new methodology to quantify the systematic biases arising from significant diffuse contributions. 
We applied this approach in the analysis of EGRET data
to study gamma-ray emission from pulsars for energies above 100 MeV.

Pulsars represent astrophysical laboratories for extreme conditions.
Their properties such as densities, temperatures, velocities, electric potentials, 
and magnetic fields associated with these spinning neutron stars give rise to high-energy 
emission through a variety of mechanisms.
Before the launch of CGRO in 1991 only Crab pulsar (PSR B0531+21),  Vela pulsar 
(PSR B0833-45) and  Geminga (but not as pulsar in that moment) 
were known as gamma-ray sources. The instruments on CGRO have detected
a total of 7 pulsars with high significance: Crab, B1509-58, Vela, B1706-44, B1951+32, Geminga, and
B1055-52. The weakest (PSR B1951+32) has a statistical probability of occurring by chance of $\sim
10^{-9}$.  Not all seven are seen at highest energies: PSR B1509-58 is seen only up 
to 10 MeV by COMPTEL and not at 100 MeV by EGRET. 
Sensivity for an individual PSR detection is greatly enhanced once studied phase-coherent.
The six seen by EGRET all show a double peak in their light curve. In
addition to the six high confidence pulsar detection above 100 MeV, three additional radio
pulsars may have been seen by EGRET: B1046-58, B0656+14, J0218+4232. 
These three all have chance probabilities about 5 orders of
magnitude less convincing than PSR B1951+32.

More than 1500 radio pulsars are known and it can be expected that
this number will continue to grow as more refined detection equipment is used 
and spatial coverage is expanded. 
The gamma ray pulsars can be compared to other pulsars
in term of the derived physical parameters. Figure 1 (from reference
\cite{Thompson2003}) displays the
distribution of observed radio pulsars in a period-period-derivative diagram
derived from Australia Telescope National Facility (ATNF) Pulsar Catalogue \cite{ATNF} \footnote{Figure 1 is from 2003, so  pulsars discovered since them are not included.}  Lines
indicating the "rotational" age of the pulsars and their dipole
field strength are also shown as well as the open field line voltage. 
The gamma-ray pulsars are shown as
squares (large dark boxes: seven high-confidence gamma-ray pulsars;
large light boxes: three lower-confidence gamma-ray pulsars). 
Gamma ray pulsars tend to be concentrated in region with high magnetic field (shown by dashed
lines), relatively young ages (shown by solid lines) and their open field line voltage is high
compared to most pulsars (dotted lines).
\begin{figure}
\centering
\includegraphics[width=0.3\textwidth]{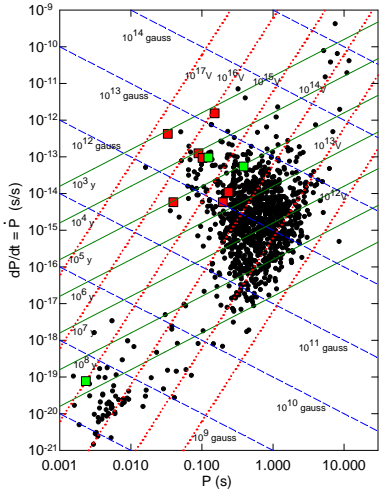}
\caption{Distribution of observed radio pulsars in a period-period-derivative 
diagram derived from ATNF Pulsar Catalogue \cite{ATNF}. The figure is from reference
\cite{Thompson2003}. See text.}
\label{fig:1}
\end{figure}
Efforts to search for additional pulsars in EGRET data have been unsuccessful due to 
limited statistics. In order to apply the stacking technique in the study of gamma-ray 
pulsars we created subclasses of pulsars using ATNF pulsar survey after sorting them with 
different criteria:
Surface magnetic flux density ($F_{B}=3.2\times10^{19}(P/ \dot{P})^{1/2}$, where $P$ is the pulsar's period),  
best estimate of the pulsar distance ($D$), Spin down energy loss rate ($\dot{E}$), Energy flux at 
the Sun ($F_{E}/D^2$), Spin down age ($\tau=P/(2\dot{P})$). The classes were chosen
accordingly on what is already known about the gamma-ray pulsars that have
been detected. We excluded from each list of subclasses those pulsars with $|b|>30^{\circ}$ 
and the detected EGRET pulsars: Crab, J0633$+$1746,
Vela, PSRB1055$-$52, PSRJ1706$-$44, PSRB1951$+$32 (see for example:
\cite{FMNT,Thompson1998,Thompson2003} and also J2229$+$6114 \cite{TDNR}).
Those high-gamma-ray flux pulsars were excluded because they inmediately 
determine the complete stacking problem.
We first analyzed pulsars individually using the standard EGRET software.
After that, for each class or subclass we have generated stacked maps
containing N pulsars, with N=2, 4, 6, ..., 50. For each stacked map so
generated, we have then determined the flux, flux error, upper limit,
$TS$, $g_{mul}$, $g_{bias}$, $g_{mul}/g_{bias}$ in the center of the
maps and $g_{mul}/g_{bias}$ averaged over a $6^{\circ} \times 6^{\circ}$ box, 
approximately the size of EGRET's PSF for energies $>$ 100 MeV.
\label{stacking_pulsars}
\section{Results}
Figure 2 shows some of the results of our study for all the subclasses of pulsars 
investigated: $\dot{E}$, $F_{B}$, $\tau$, $D$ and $F_{E}/D^2$.
The plots represent the $TS$ obtained (left-axis) and $g_{mul}/g_{bias}$ averaged 
over $6^{\circ} \times 6^{\circ}$ box (right-axis) versus the number of stacked maps.
We did not find any signal in the $F_{E}/D^2$ class.
$F_{B}$  and $\tau$ class both have similar behavior with a peak that appears at high N values 
in the sample. On the other hand, $\dot{E}$ class has a peak that is dominated from a few sources. 
The $D$ class is more spread; but a  tendency to show high values of TS is apparent perhaps for the more nearby sources, then
fading away when distances of about 1 kpc and larger are reached. 
\begin{figure}
\centering
\includegraphics[width=0.27\textwidth]{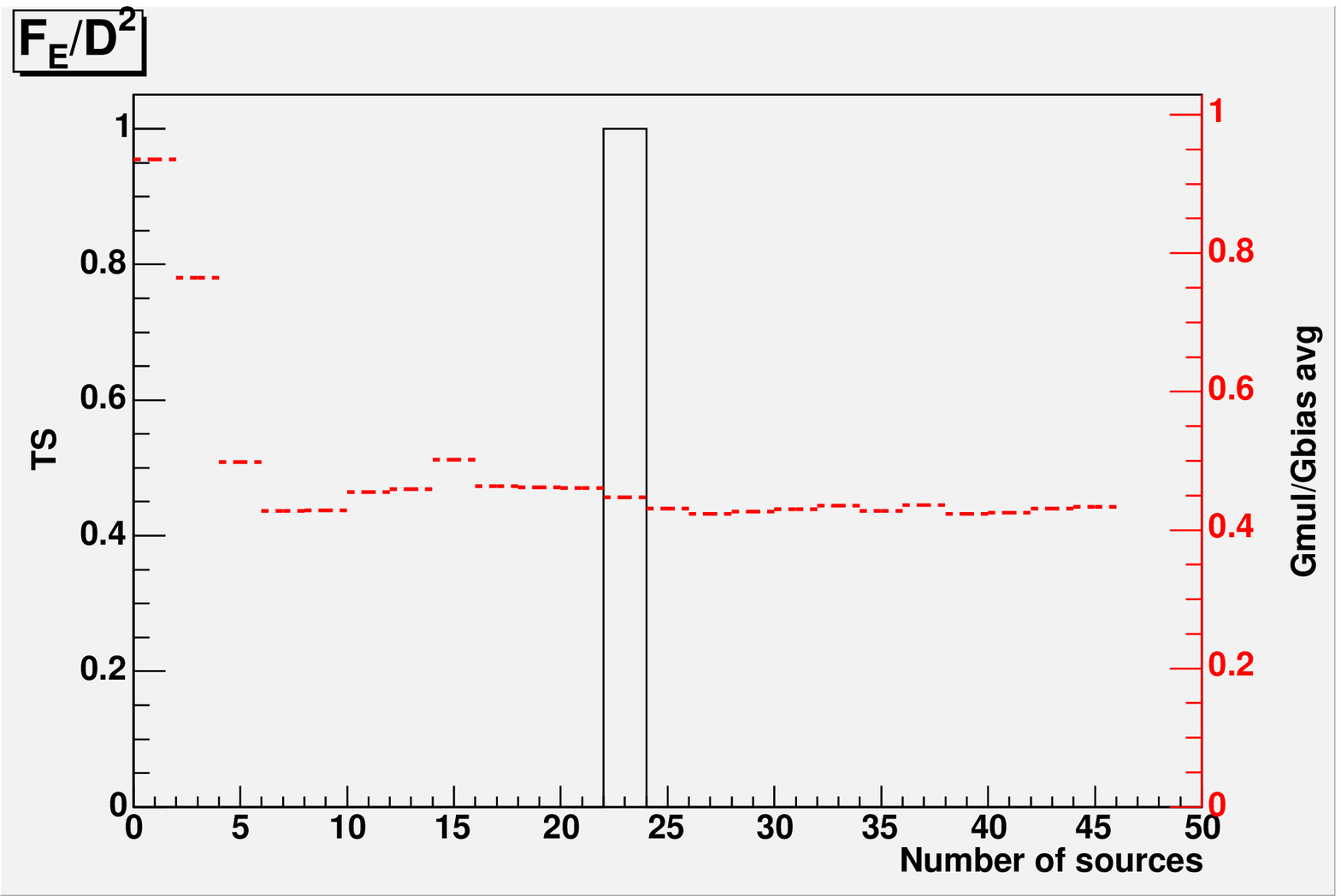}
\includegraphics[width=0.27\textwidth]{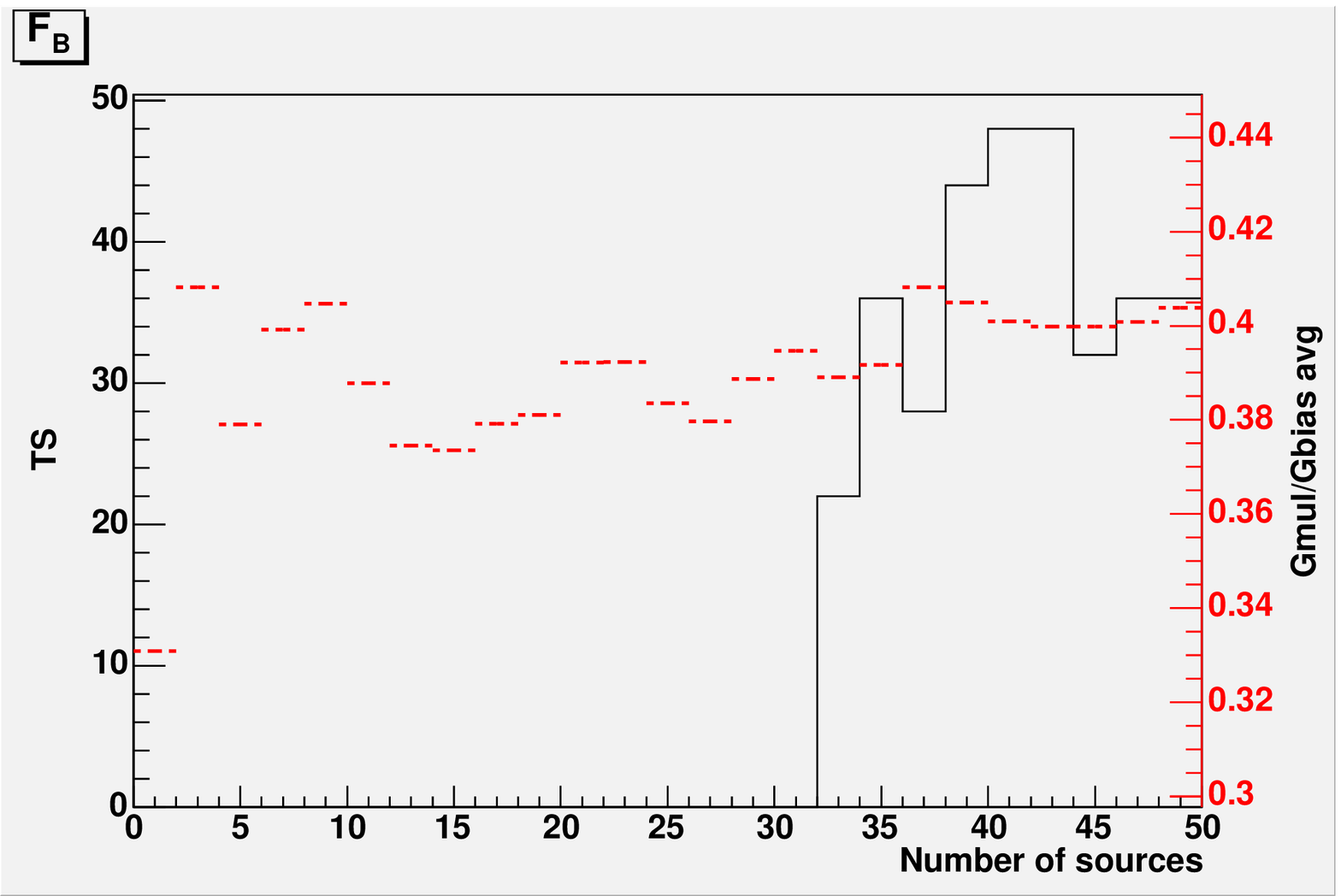}
\includegraphics[width=0.27\textwidth]{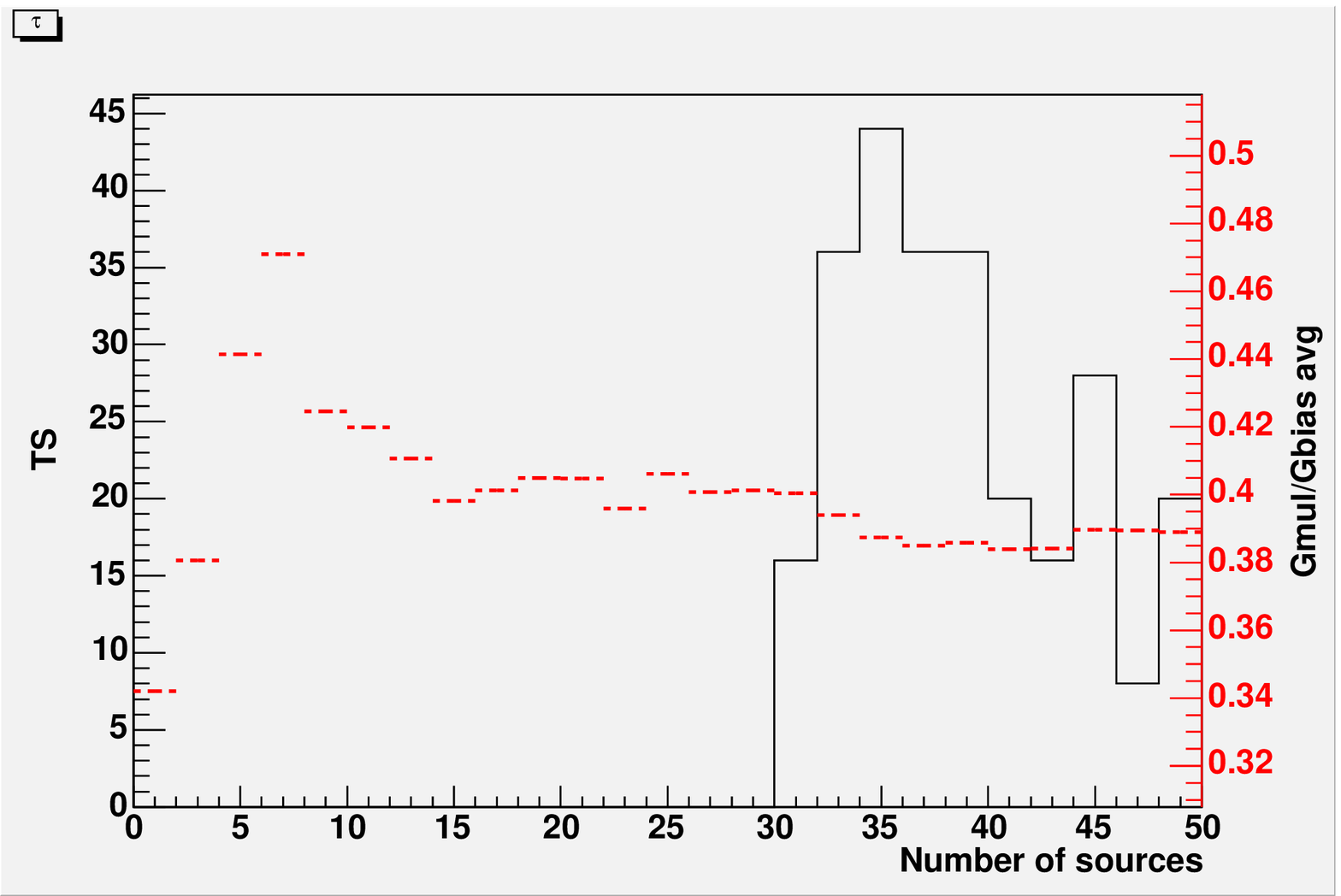}
\includegraphics[width=0.27\textwidth]{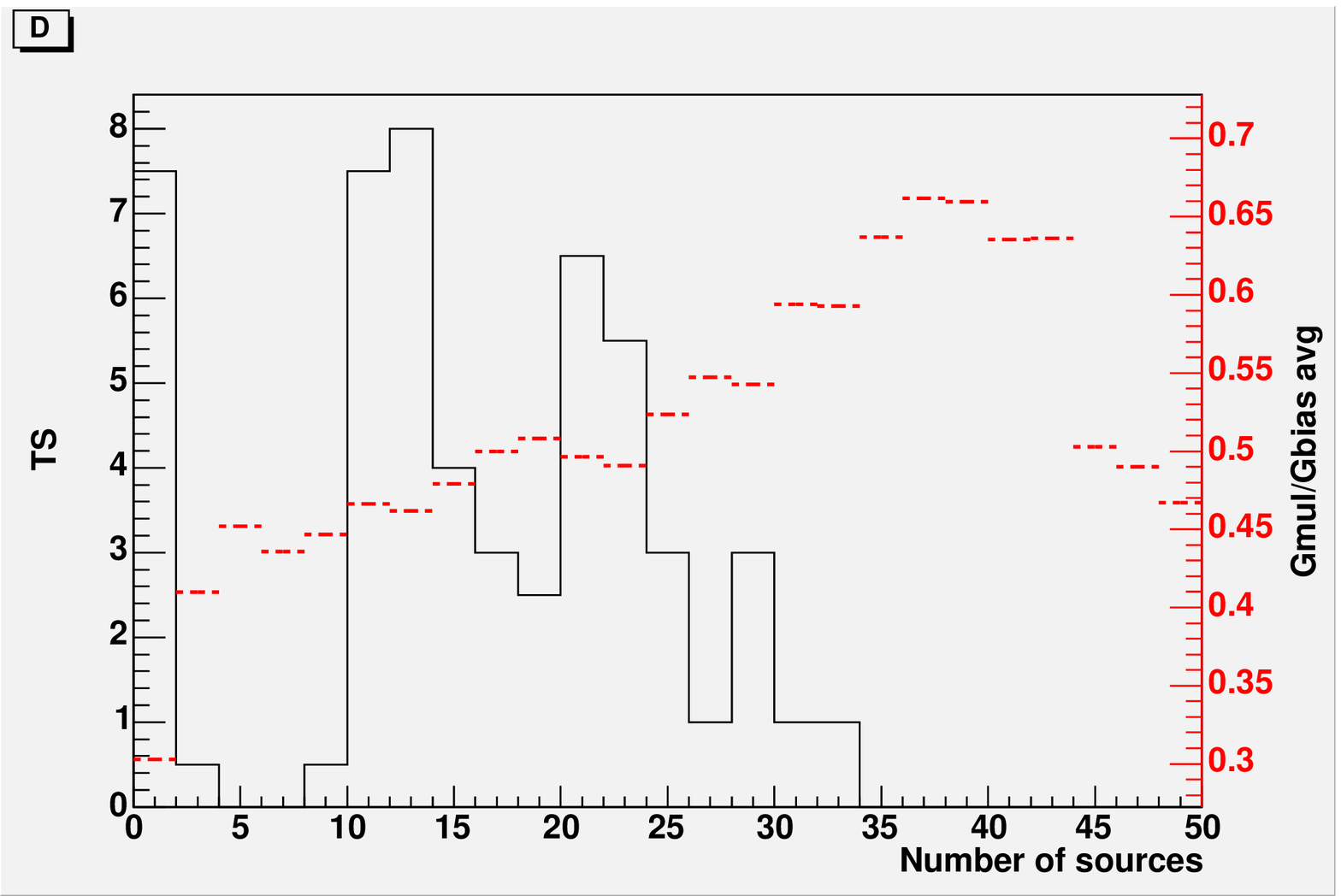}
\includegraphics[width=0.27\textwidth]{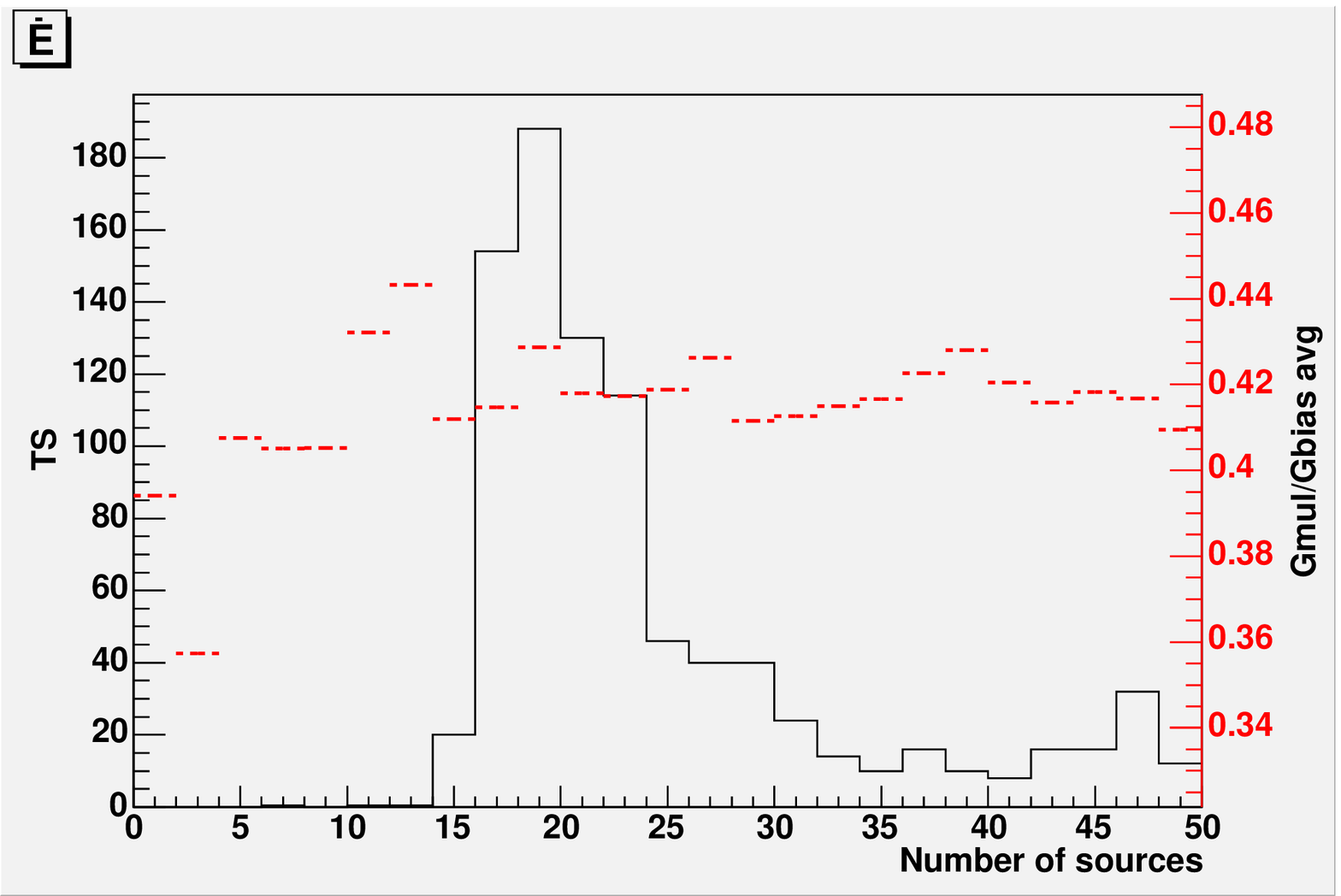}
\caption{$TS$($g_{mul}/g_{bias}$) versus number of sources added
left-axis(right-axis) for different subclasses of pulsars
investigated: $\dot{E}$, $F_{B}$, $\tau$, $D$ and $F_{E}/D^2$.  $TS$, solid lines; $g_{mul}/g_{bias}$, dashed lines. 
See text.}
\label{fig:2}
\end{figure}

During stacking, we keep continuously track on the individual contribution from the diffuse 
emission model. By comparing the individual contribution and evolution of the diffuse emission over the growing numbers of sources in the stacking sample, we can immediately judge if 
a change in $TS$ is due to a newly added source or rather to a odd diffuse emission 
value/problematic treatment during the stacking. There are several cases to distinguish 
already: (a) steadily accumulating $g_{mul}/g_{bias}$ as in Figure 2 d (D), which points 
towards an steadly increasing dominance of the diffuse emission, thus diminishing the 
chance for determine equal $TS$ when adding more sources. (b) constant 
$g_{mul}/g_{bias}$ as in Figure 2 a, b, c, e - which assures that no glitches in the 
diffuse model have an impact on the outcome of the stacking result. 
Initial jumps at the beginning of the stacking are compensated in the average after $\sim$ 5-6 sources in sample.
\label{Results}
\section{Monte Carlo Simulations}
In order to understand the results obtained above we ran
Monte Carlo Simulation creating fictitious objects with zero gamma-ray flux
at random sky positions with $|b|<30^{\circ}$, transforming and
co-adding the maps, then analyzing the stacked map using the same methods as for real
objects. 1000 simulations were performed for 2, 10,
20, 30, and 40 objects added with zero gamma-ray flux.
Examples of the results obtained are shown in Figure 3, where the $TS$
cumulative distribution is plotted for 2, 20 and 40 sources added.
\begin{figure}
\centering
\includegraphics[width=0.33\textwidth]{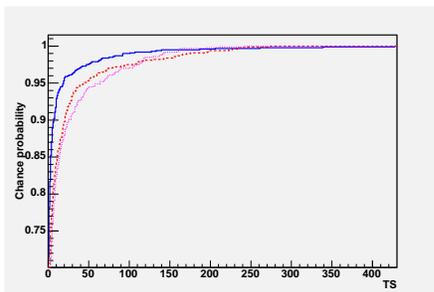}
\caption{Monte Carlo Simulations: Chance probability of $TS$.
2 (solid line), 20 (dashed line) and 40 (dot line) sources added
with zero gamma-ray flux. See text.}
\label{fig:3}
\end{figure}
Accordingly to our results there is no more than 3 \% chance of
obtaining $TS>$ 50 if 2 sources are added; and no more than 6\% if 10 (not shown in Figure 3), 20, 30 (not shown in Figure 3), or 40  objects with zero 
flux are added randomly in the GP. 

\label{simulations}
\section{Discussion}
In this paper a new technique to do source population studies using the stacking 
method in the GP is presented. Different subclasses of pulsars were investigated. 
Among all orderings of pulsars, the one sorted by distance ($D$) is the most 
promising, following the behavior that we expected previous to our study: $TS$ 
signal fading away when the diffuse contribution grows. The fact that the ordering by 
$F_{E}/D^2$ does not  appear to show a signal in the stacking is surprising 
and will be investigated for systematics in subsequent studies. 
The $F_{B}$ and $\tau$ classes start to show a signal for quite large number of pulsars,
 which may be due to the fact that 
only individual sources affect the stacking 
. 
 We anticipate that the technique explained
 in this paper will have application 
in the study of Galactic sources (not only for PSRs) in GLAST era. 
\label{discussion}
\section*{Acknowledgments}
DFT has been supported by Ministerio de Educaci\'on y Ciencia (Spain) 
under grant AYA-2006-0530, as well as by the Guggenheim Foundation.


\begin{thebibliography}{3}
%
\bibitem{Torres2003}
Torres et al.: Supernova remnants and gamma-ray sources. Phys Report 382, 303 (2003)
%
\bibitem{RBT}
Romero G., Benaglia P., \& Torres D. F : Unidentified 3EG gamma-ray sources at low galactic latitudes. A\&A 348, 868 (1999) 
%
\bibitem{CHB}
Cillis A. N., Hartman R. C. \& Bertsch D. L.: Stacking Searches for Gamma-Ray Emission above 100 MeV from Radio and Seyfert Galaxies. ApJ 601, 142 (2004)
%
\bibitem{Hartman}
Hartman R. C. et al.: The Third EGRET Catalog of High-Energy Gamma-Ray Sources. ApJS 123, 79 (1999)
%
\bibitem{Hunter}
Hunter S. D. et al.: EGRET Observations of the Diffuse Gamma-Ray Emission from the Galactic Plane. ApJ 481, 205 (1997)
%
\bibitem{MHR}
Mattox J. R, Hartman R. C. \& Reimer O.: A Quantitative Evaluation of Potential Radio Identifications for 3EG EGRET Sources. ApJS 135, 155 (2001)
%
\bibitem{Mattox}
Mattox J. R. et al.: The Likelihood Analysis of EGRET Data. ApJ 461, 396 (1996)
%
\bibitem{CTR}
Cillis A. N., Torres D. F. \& Reimer O.: EGRET Upper Limits and Stacking Searches of Gamma-Ray Observations of Luminous and Ultraluminous Infrared Galaxies. ApJ 621, 139 (2005)
%
\bibitem{RPSM}
Reimer O., Pohl M., Sreekumar P. \& Mattox J. R.: EGRET Upper Limits on the High-Energy Gamma-Ray Emission of Galaxy Clusters. ApJ 588, 155 (2003)
%
\bibitem{Thompson2003}
Thompson D. J.:  Gamma Ray Pulsars: Multiwavelength Observations. astro-ph/0312272 (2003)
%
\bibitem{ATNF}
Manchester, R. N., Hobbs, G. B., Teoh, A. \& Hobbs, M.: The Australia Telescope National Facility Pulsar Catalogue. Astron. J., 129, 1993-2006 (2005) \\
    http:\/\/www.atnf.csiro.au\/research\/pulsar\/psrcat.
%
\bibitem{FMNT}
Fierro J. M., Michelson P. F., Nolan P. L., \& Thompson D. J.: Phase-resolved Studies of the High-Energy Gamma-Ray Emission from the Crab, Geminga, and VELA Pulsars. ApJ 494, 734 (1998)
%
\bibitem{Thompson1998}
Thompson D. J. et al.: Gamma Radiation from PSR B1055-52. ApJ 516, 297 (1999)
%
\bibitem{TDNR}
Thompson D. J., Digel S. W.,  Nolan P. L., \& Reimer O.: Neutron Stars in Supernova Remnants. 
ASP Conference Proceedings, eds P. O. Slane and B. M. Gaensler (2001), 
astro-ph/0112518v1
%
\end{thebibliography}
\end{document}